\documentclass{aa}

\usepackage{graphicx}
\usepackage{natbib}
\usepackage{amsmath}
\usepackage{amssymb}
\usepackage{txfonts}

\graphicspath{ {./pic/} } 

\bibpunct{(}{)}{;}{a}{}{,}

\title{\textsl{XMM-Newton} Observation of the Anomalous X-Ray Pulsar
  4U~0142+61}

\author{E.~G\"ohler\inst{1} \and J.~Wilms\inst{2} \and R.~Staubert\inst{1}}

\institute{Institut f\"ur Astronomie and Astrophysik, Abt.\ Astronomie,
  University of T\"ubingen, 72076 T\"ubingen, Germany
  \and 
  Department of Physics, University of Warwick, Coventry CV4 7AL,
  United Kingdom
  }

\date{Received 29 October 2004 /Accepted 2 October 2004}

\abstract{ We present results of an observation of the anomalous X-ray
  pulsar 4U~0142+61 with the EPIC cameras on \textsl{XMM-Newton}
  performed on 2003 January 24.  The pulse phase averaged spectrum can
  be best described by the sum of a black body with a temperature of
  $kT_\text{BB} = 0.395(5)\,\text{keV}$ and a power law with photon
  index $\Gamma=3.62(5)$. The unabsorbed 2--10\,keV flux is
  $\sim$$7.2\cdot 10^{-11}\text{erg}\,\text{s}^{-1}\,\text{cm}^{-2}$.
  No evidence for additional spectral features such as cyclotron lines
  is present.  These results are consistent with those from an earlier
  \textsl{Chandra} observation in 2001 May.  Phase resolved
  spectroscopy over the $8.6882(2)$\,s period (MJD 52663.93) shows
  clear variations with pulse phase of $\Gamma$, while $kT_\text{BB}$
  shows a small variation of $\sim 12\%$. We confirm earlier conclusions
  by \citet{oezel:01b} that the emission from AXPs is more likely to
  originate from neutron stars with one single and magnetically active
  region on the neutron star. The significantly different behavior of
  the soft and hard spectral components with pulse phase, however,
  cannot be fully reconciled with the present magnetar emission
  models.  }

\begin{document}
\maketitle

\section{Introduction}
4U 0142+61 is a member of the small group of anomalous X-Ray pulsars
(AXPs) which are characterized by spin periods in the range of
8--12\,sec, associated with a relatively stable long term spin down
\citep{gavriil02:_long_term_rxte}, very soft X-ray spectra well
modeled by blackbody plus power law components with $\Gamma\sim 3$, a
X-ray luminosity of
$\sim$$10^{34}$--$10^{36}\text{erg}\,\text{s}^{-1}$, low in comparison
to High Mass X-Ray Binaries (HMXB) \citep{israel:02_obs_axp}, and very
faint optical counterparts.  It is generally assumed that these
objects are isolated neutron stars as no binary companion for any of
these objects has been found up to now.

The X-ray luminosity of AXPs exceeds the energy available from the
spin down of the neutron star, so some additional energy source is
needed \citep{paradijs:95:6s_x}.  There are mainly two classes of
models trying to explain the source of the missing energy. One class
proposes that the AXPs are powered by accretion, either from debris of
a disrupted HMXB after a common envelope phase
\citep{paradijs:95:6s_x} or from a disk formed by the fallback
material from a supernova explosion \citep{chatterjee:00_axp_p_l_age}.
The other class of models assumes a very strong magnetic field which
can explain the spin down as magnetic dipole radiation while the X-ray
luminosity is supplied by either the decay of the magnetic field
\citep{thompson:95:magnetar_i} or the cooling of the neutron star
\citep{heyl:97_ns_cooling}.

The brightest of the small sample of anomalous X-ray pulsars is
4U~0142+61, a source detected by \textsl{Uhuru}, which had its nature
established during an \textsl{EXOSAT} campaign in which a $8.7\,$s
periodic variation was found \citep{israel94:_discovery_pulsation}.
Subsequent X-ray observations manifested features usually attributed
to AXPs such as a spectrum which is best fit by a 0.386(5)\,keV black
body plus a power law with photon index $\Gamma=3.67(9)$,
\cite{white96:_spect_asca}. Long term period changes were discovered
in subsequent \textsl{ASCA} and \textsl{RXTE} observations
\citep{paul00:_study_long_term_stabil_asca,gavriil02:_long_term_rxte},
with the latter revealing a period of $P=8.68832877(3)$\,s and a
period change of $\dot{P}=2.02(25)\times
10^{-12}\,\text{s}\,\text{s}^{-1}$.  Finally, an optical counterpart
was detected which showed strong periodic variation \citep[][see also
\citealt{hulleman:04}]{kern02:_optic_x}.  A recent observation of
4U~0142+61 by \textsl{Chandra} confirmed the association of this
optical counterpart with the X-ray source and also resulted in a
further improvement of the X-ray spectral results
\citep{patel03:_chandra_obser_acis,juett02:_chand_high_resol_spectra}.

In this Paper we report an observation of 4U~0142+61 with
\textsl{XMM-Newton}.  In Sect.~\ref{sect:observation} we describe the
observation and data reduction. Lightcurve analysis and pulse phase
averaged spectral analysis, presented in
Sect.~\ref{sect:timing_analysis} and
Sect.~\ref{sect:averaged_spectra}, confirm earlier \textsl{Chandra}
and \textsl{RXTE} results.  Phase resolved spectroscopy shows a
variation of two different spectral components which are presented in
Sect.~\ref{sect:phase_resolved_spectra} and discussed in
Sect.~\ref{sect:discussion}.

\section{Observations  with \textsl{XMM-Newton}}
\label{sect:observation}
The observation of 4U~0142+61 with \textsl{XMM-Newton} was performed
on 2003 January 24 (\textsl{XMM-Newton} Revolution 573).  This
observation has an exposure time of $\sim$4.2\,ksec with the EPIC-pn
camera in small window mode and $\sim$5.8\,ksec with both EPIC-MOS
cameras in fast timing mode.  The EPIC-pn detector is a
back-illuminated charge coupled device (CCD) using $64\times 64$
pixels in small window mode, covering a $\sim$$4\arcmin\times
4\arcmin$ area \citep{strueder01:pn_camera}.  The two MOS cameras are
front-illuminated CCDs detectors \citep{turner01:mos_camera}.  In
their timing mode, spatial information is preserved only along the
$x$-axis of the detector, but events are registered with a higher time
resolution than in the EPIC-MOS imaging modes.

Standard data processing was performed with the \textsl{XMM-Newton}
Scientific Analysis Software, version 5.4.1.  For the EPIC-pn the
source region was taken to be a circle of radius $\sim 100\arcsec$
around the point source. The background was measured in two
rectangular regions arranged at the corners of the readout window.
For the MOS cameras, events in the raw $x$ interval $280\ldots 330$
were chosen for source events, background events were taken from the
intervals $260\ldots 279$ and $331\ldots 345$. No background flares
were present during this observation. We checked the EPIC-pn data for
evidence for pileup and found no indication for pattern pileup or
pileup in diagonal pixel patterns at energies above 0.3\,keV.  For the
MOS1/2 data pattern pileup is negligible in timing mode
\citep{ehle03:_xmm_user_handbook}.

We note that a second earlier \textsl{XMM-Newton} observation of
4U~0142+61 exists, which was performed on 2002 February 13 (revolution
399).  For this 3.4\,ksec long observation, only data from the EPIC-pn
cameras are available (the MOS cameras were switched off) and the
background was a factor of 10 higher than in the Rev.~573 observation,
such that we decided to not use these data here.

The position of 4U 0142+61 was estimated by fitting a Gaussian to the
point-spread function of the EPIC-pn data in both observations.  The
position obtained is
$\alpha_{\text{J}2000.0}=01^\text{h}\,46^\text{m}\,22\farcs 3$,
$\delta_{\text{J}2000.0}=61\degr 45\arcmin 02\farcs 5$, which is
consistent within the XMM $2\farcs 2$ error circle with the more
accurate \textsl{Chandra} position of $\alpha=01^{\rm h}\,46^{\rm
  m}\,22\farcs 42$, $\delta=61\degr 45\arcmin 02\farcs 8$
\citep[positional uncertainty $\sim 0\farcs
7$;][]{patel03:_chandra_obser_acis}.

\section{Timing analysis}
\label{sect:timing_analysis}
For all instruments the lightcurve was searched for periods in the
$8.6\,$s domain using epoch folding
\citep{leahy83:pulsed_emission_epfold}.  Combining the results of the
period search from all instruments of Rev.~573 (Epoch: MJD 52663.93)
we found a period of $P=8.6882(2)\,$s.  This period agrees with the
RXTE data of \citet{gavriil02:_long_term_rxte} which predict
$P=8.68849(1)\,$s, assuming a constant $\dot{P}=2.02(25)\times
10^{-12}\,\text{s}\,\text{s}^{-1}$.

We estimate the uncertainty of the \textsl{XMM-Newton} period using a
Monte-Carlo simulation: A synthetic lightcurve was constructed by
taking the time stamps as in the measured light curve and associate
them with the flux of the folded pulse profile at the corresponding
phase (for the best period and ephemeris found from the observed
data).  From this template 1000 simulated lightcurves were constructed
assuming the mean pulse profile and then randomizing the flux values
using the appropriate Poisson distribution.  For all simulated
lightcurves the best periods were determined.  The standard deviation
of the distribution of those periods was taken as the uncertainty of
the period found in the observed data.

\begin{figure}
\resizebox{\hsize}{!}{\includegraphics{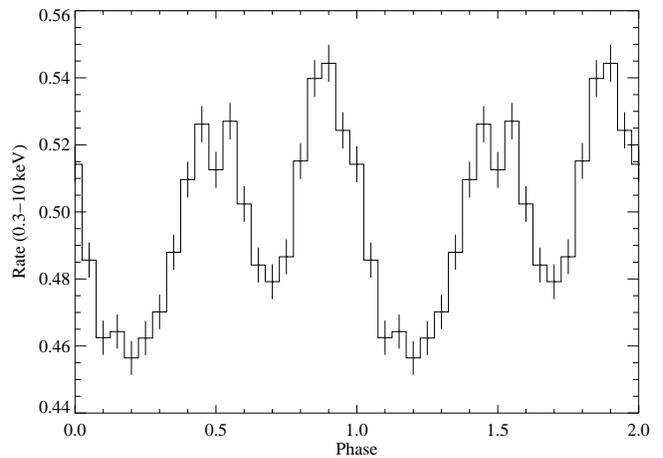}}
\caption{\label{total_profile} Pulse profile of 4U~0142+61 in the
  energy range 0.3--10\,keV.}
\end{figure}

For further timing and profile analysis we used a period of
$8.68823\,$s.  To compute pulse profiles all single events from
all three EPIC detectors are added. Multiple events were not taken
into account.
The pulse profile in the energy range 0.3--10\,keV is shown in
Fig.~\ref{total_profile}.  Its peak-to-peak pulsed fraction, defined
as $(F_\text{max}-F_\text{min})/(F_\text{max}+F_\text{min})$, is
$7.7\%\pm 0.9$\%.

\begin{figure}
\resizebox{\hsize}{!}{\includegraphics{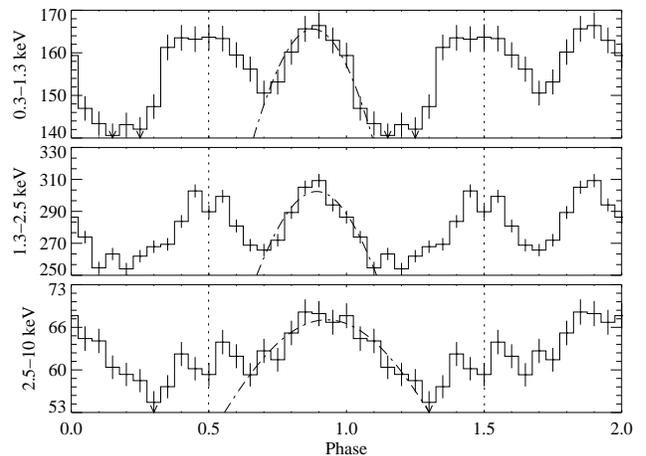}}
\caption{\label{e_resolved_profile}
  Profile of the combined lightcurves in the energy ranges
  0.3--1.3\,keV, 1.3--2.5\,keV, and 2.5--10\,keV. The peak-to-peak
  pulse fractions are: $8.6\%\pm 1.3\%$ (0.3--1.3\,keV), $9.1\%\pm
  0.9\%$ (1.3--2.5\,keV), and $9.6\%\pm 2.0\%$ (2.5--10\,keV).  The
  dash dotted line is the result of the fit of a gaussian distribution
  to the peak at phase $\sim 0.9$.  Here all data from EPIC-pn and
  MOS-1/2 from Rev.~573 were used.}
\end{figure}

In Fig.~\ref{e_resolved_profile} we display the pulse profile and give
the peak-to-peak pulse fractions for several energy bands. As was
first noted by \cite{white96:_spect_asca} the pulse is double peaked
at low energies. At higher energies the main pulse remains as a broad
single peak centered at phase $\sim$0.9, while the secondary peak
weakens and changes into a double peaked feature.

In our data we observe a slight phase shift of the main peak. Fitting
a Gaussian distribution to this peak yields a centered phase value of
0.878(9) (0.3--1.3\,keV), 0.891(5) (1.3--2.5\,keV) and 0.92(1)
(2.5-10\,keV). The peak-to-peak pulsed fraction is not significantly
energy dependent, with measured pulsed fractions of $8.6\%\pm 1.3\%$
(0.3--1.3\,keV), $9.1\%\pm 0.9\%$ (1.3--2.5\,keV), and $9.6\%\pm
2.0\%$ (2.5--10\,keV).

\section{Spectra}
\label{sect:averaged_spectra}
For spectral analysis we used the single event data from the EPIC-pn
camera and the MOS cameras of Rev.~573.  Simultaneous fits were made
with XSPEC version 11.2.0bl in the energy ranges of 0.5--9.5\,keV (pn)
and 0.5--8\,keV (MOS), using a multiplicative constant to take into
account the flux calibration uncertainty of the MOS cameras with
respect to the EPIC-pn.  To allow $\chi^2$ analysis the channels were
rebinned to contain at least 50\,counts per bin.

Several models were used to fit the phase averaged spectra: single
component models (blackbody, bremsstrahlung, power law) and
combinations of those, including the classical combination of a black
body component plus a power law.  All models include absorption due to
the interstellar medium with hydrogen equivalent column density
$N_\text{H}$ (using the XSPEC model \textit{phabs}).  Fit results of
the different models are shown in Table~\ref{table:spec_results}.

\begin{table*} 
\caption{Phase averaged spectral fits%
  \label{table:spec_results}. Best fit parameters for an
  absorbed power law (PL) and a bremsstrahlung (BREMS) continuum, with and
  without an additional blackbody (BB) component. The parameters are
  the equivalent hydrogen column density, $N_\text{H}$, the power-law photon
  index, $\Gamma$, and its normalization at 1\,keV in units of
  $10^{-1}\,\text{Photons}\,\text{keV}^{-1}\text{cm}^{-2}\text{s}^{-1}$,  
  the temperature of the bremsstrahlung spectrum, $kT_\text{brems}$,
  and its 2--10\,keV component flux contribution, $F_\text{brems}$, in 
  $10^{-11}\text{erg}\,\text{s}^{-1}\,\text{cm}^{-2}$ for the EPIC-pn camera,
 the temperature of the black body, $kT_\text{bb}$,
  and the radius of the black body, $R_\text{bb}$, assuming a distance
  of 10\,kpc. MOS$\,$1/2 spectra are scaled with constants
  $S_\text{MOS1}$ and $S_\text{MOS2}$ with respect to the EPIC-pn. All
  error bars are at the 90\% level for one interesting parameter  and
  expressed  in parentheses in units of the last digit shown.
}
\begin{center}
\begin{tabular}{*{11}{l}}\hline\hline
      & $N_\text{H}$      
      & $\Gamma$          & $N_\Gamma$ 
      & $kT_\text{brems}$ & $F_\text{brems}$
      & $kT_\text{bb}$    & $R_\text{bb}$       
      & $S_\text{MOS1}$   & $S_\text{MOS2}$
      & $\chi^2/\text{DOF}$\\ 
Model & $(10^{22} \text{cm}^{-2})$
      &                   &
      & (keV)             & 
      & (keV)             &  (km)     
      &                   &
      &  \\\hline
PL    & $1.2(5)$          
      & $3.9(1)$          &   $0.39(4)$
      &   ---             &   ---         
      &   ---             &   ---         
      &  0.84(4)          &  0.85(4)
      & $3039/1144$\\    
BREMS & $0.82(4)$         
      &  ---              &   ---
      &  $1.09(5)$        &  6.79(2)
      &   ---             &   ---      
      &  0.85(4)          &  0.85(4)
      & $1721/1144$\\
BB+BREMS&$0.68(8)$
      & ---               &   --- 
      &  $1.7(4)$         & $4.5(2)$
      &  $0.35(3)$        & $835(27) $
      &  0.85(4)          & 0.85(4)
      &  $1299/1142$ \\
BB+PL & $0.96(2)$
      & $3.62(5)$         & $0.155(8)$
      &   ---             &   ---
      & $0.395(5)$        &  428(40)     
      & 0.85(4)           &  0.85(4)
      & $1367/1142$\\\hline
\end{tabular}
\end{center}
\end{table*}

\begin{figure}
\resizebox{\hsize}{!}{\includegraphics{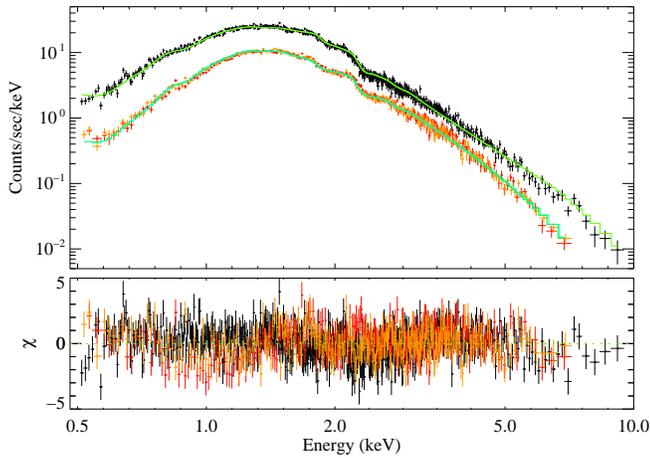}}
\caption{\label{figure:phaseaveraged_bb_powerlaw_fit}
Spectrum and residuals for the blackbody plus power law model.}
\end{figure}

Single component models gave unacceptable results. The best fit is
obtained for the black body plus bremsstrahlung model.  The blackbody
plus power law model (Fig.~\ref{figure:phaseaveraged_bb_powerlaw_fit})
gave almost the same $\chi^2$ but had a smoother residual at higher
energies, and we will use this model in our further discussion.  The
residuals of this model show no indication for any line in the energy
band covered by our observation.  Using this model the unabsorbed
2--10\,keV flux is
$7.02\times10^{-11}\,\text{erg}\,\text{s}^{-1}\,\text{cm}^{-2}$.  No
further spectral features such as cyclotron lines were found in the
spectral range of 0.5--10\,keV.

Extending the fit to energies below 0.5\,keV for the EPIC-pn data
alone and refitting yields a deviation of $\sim$30\% between the data
and the model in 0.3--0.5\,keV band.  This deviation is larger than
allowed by the uncertainties of the calibration
\citep{kirsch:03:tn_18}.  Phase resolved spectroscopy shows this
excess of the model in all phases (see
Sect.~\ref{sect:phase_resolved_spectra}), i.e., the excess flux is not
caused by a possible continuum variability with pulse phase.  The
feature also is present for different absorption models and abundances
and is also present in the double event spectrum of the EPIC-pn
camera.  We assume that the residual either comes from a higher
galactic absorption than the absorption model or from unresolved
calibration issues in this energy band. For the remainder of this
paper, therefore, we will restrict ourselves to the energy band above
0.5\,keV.

\section{Phase resolved spectroscopy}
\label{sect:phase_resolved_spectra}
\begin{figure}
\resizebox{\hsize}{!}{\includegraphics{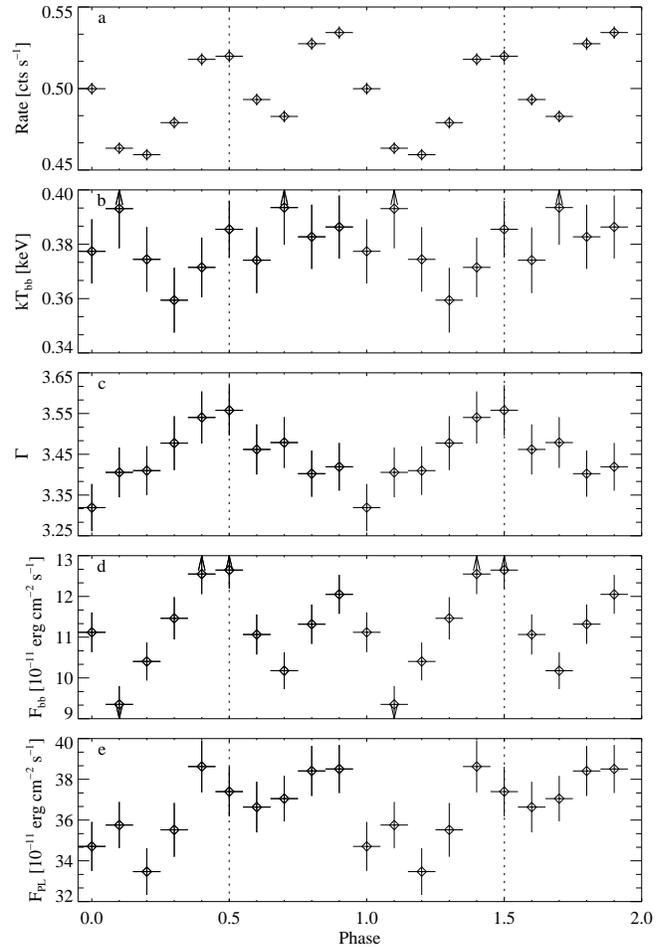}}
\caption{\label{figure:phaseres} Results of phase resolved
  spectroscopy for a blackbody plus power law model. Shown are
  \textbf{a} total EPIC count rate, \textbf{b} blackbody temperature,
  \textbf{c} photon index, \textbf{d} blackbody flux, and \textbf{e}
  flux of the power law.  The 0.5--10\,keV fluxes are unabsorbed
  fluxes and are in units of $10^{-11}\,\rm erg\,cm^{-2}\,s^{-1}$. }
\end{figure} 

To search for spectral variations with pulse phase we split the data
into ten phase bins, rebin the spectra to contain at least 25 counts
per bin, and model the EPIC data of each phase bin with a black body
plus power law model. Fig.~\ref{figure:phaseres}a shows the resulting
pulse profile.

In a first step, all parameters were allowed to freely vary with pulse
phase.  We found that the column density $N_\text{H}$ did not show
significant variation.  To test this constancy we fit the subset of
phase bins with the highest flux, constraining $N_\text{H}$ to be the
same for these bins.  In a second iteration, $N_\text{H}$ was
allowed to vary with pulse phase.  The $F$-test shows a probability of
71\% that $N_\text{H}$ does not vary.  Therefore we considered
$N_\text{H}$ to be a constant parameter with a value of $N_\text{H}=
0.91\times 10^{22}\,\text{cm}^{-2}$ in our subsequent analysis.

The results of our spectral analysis are shown in
Figs.~\ref{figure:phaseres}b and c as a function of pulse phase. The
blackbody temperature is virtually constant, while the photon index
shows a clear pulse phase dependence.  Using the unabsorbed spectrum,
we furthermore derive the contribution of the blackbody and the power
law component to the total 0.5--10\,keV flux as a function of pulse
phase.  The result is shown in Figs.~\ref{figure:phaseres}d and~e.
Uncertainties are computed using a Monte-Carlo approach randomizing
the input data sets.  We compare these results with AXP emission
models in the following section.

\section{Discussion}
\label{sect:discussion}
In this paper we have shown that the comparatively short observation
of 4U~0142+61 with \textsl{XMM-Newton} confirms the spectral and
timing parameters from earlier observations. The best fit models are
the standard black body plus power law and the black body plus
bremsstrahlung model.  The period agrees well with the high precision
value derived from the long term timing campaign with RXTE
\citep{gavriil02:_long_term_rxte}.

The flux of $7.02\times10^{-11}{\rm erg\,sec^{-1}\,cm^{-2}}$ is
slightly lower than the flux reported in the \textsl{ASCA}
observations of \cite{paul00:_study_long_term_stabil_asca}, ranging
from $9\ldots16\times10^{-11}{\rm erg\,sec^{-1}\,cm^{-2}}$.  We
consider this discrepancy as negligible because the authors of the
\textsl{ASCA} observation note the possibility of a flux
overestimation due to limited spectral resolution.

Pulse phase resolved spectroscopy allowed us to disentangle the
different phase dependencies of the relevant parameters
(Fig.~\ref{figure:phaseres}): The blackbody temperature is constant
with pulse phase, although taken at face value there is a $\sim$12\%
variation with phase and a slightly lower temperature at phase
$\sim$0.3.  The flux of both, black body and power law, follows the
double peaked pulse profile (Fig.~\ref{figure:phaseres}a), although
but for the powerlaw flux variation the peak at phase $\sim 0.5$ is
suppressed.  This is reasonable if one takes into account the
hardening of the power law at phase $\sim 0.5$.

As has been pointed out, e.g., by \citet{oezel:01b} and
\citet{oezel:02a}, observed pulse profile variations in AXPs can be
used to test emission models for magnetars. These models predict a fan
beam plus pencil beam emission pattern and a resulting pulse profile
variation that is strongly dependent on the characteristics of the
emitting plasma. The observed pulsed fraction depends on the angle
between the magnetic field axis and the rotation axis of the neutron
star as well as on the angle between the observer's line of sight and
the rotational axis. As shown by \citet{oezel:01a}, the high pulsed
fractions of most AXPs argue for the radiation to come from a single
emitting region on the neutron star, e.g., caused by a region of
enhanced magnetic activity on the neutron star \citep{oezel:02a},
although the parameters of 4U~0142+61 are still barely compatible with
two emission regions \citep{oezel:01a}.  We can confirm this
conclusion of a single emission region from the \textsl{XMM-Newton}
data, which shows only a small change of the pulsed fraction with
energy, which is in contrast to the strong energy dependence expected
from models with two antipodal emission regions.

Using phase resolved X-ray spectroscopy, it is in principle possible
to further constrain these emission models.  In terms of the empirical
spectral model applied here, magnetar models predict that the
blackbody and powerlaw components should be coupled through resonant
Compton heating of photons scattered at cyclotron resonances
\citep[see also][]{thompson:02}. This coupling results in pulse
profiles which are predicted to have essentially the same shape over
the 1--10\,keV energy range. In our observations, this predicted
behaviour cannot be fully confirmed: Both, the energy resolved
profiles as well as the decomposition of the pulse into a blackbody
and a powerlaw component show only little correlation. For example,
Pearson's correlation coefficient, $r$, of the component fluxes of the
black body and the powerlaw is only $r=0.561$.  On the other hand,
however, the similar variation of both model components suggest that
there is a common flux contribution superimposed with an spectral
variation of the powerlaw. This would explain the decrease of the
pulse at phase $\sim 0.5$ with higher energies.

In summary, in this paper we were able to show that phase resolved
spectroscopy adds power to constrain existing models and allows in
principle to test different hypotheses describing the physical
geometry of the system. Further and longer observations and further
modeling are required, however, until these tests will yield a
conclusive and selfconsistent answer on the emission process at work.

\begin{acknowledgements}
  We acknowledge the support through DLR grants 50~OX~0002 and
  50~OG~9601. This work is based on observations obtained with 
  \textsl{XMM-Newton}, an ESA science mission with instruments and
  contributions directly funded by ESA Member States and the USA
  (NASA).
\end{acknowledgements}

\bibliographystyle{aa}
\bibliography{mnemonic,aa_abbrv,4u0142+61,axp}

\end{document}